\documentclass[preprint,sort&compress]{elsarticle}

\usepackage{amssymb}

\usepackage{subcaption}
\usepackage{natbib}

\usepackage{tabulary}
\usepackage{multirow}
\usepackage{array}

\newcolumntype{P}[1]{>{\centering\arraybackslash}p{#1}}
\newcolumntype{M}[1]{>{\centering\arraybackslash}m{#1}}

\usepackage[ruled,linesnumbered]{algorithm2e}
\usepackage[noend]{algpseudocode}

\begin{document}

\begin{frontmatter}



\title{Automated Image Analysis and Contiguity Estimation for Liquid Phase Sintered Tungsten Heavy Alloys}


\author{A. Murat Ozbayoglu}
\address{TOBB University of Economics and Technology}
\address{Department  of Computer Engineering}
\address{Ankara, 06560, Turkey}
\author{Nuri Durlu}
\address{TOBB University of Economics and Technology}
\address{Department  of Mechanical Engineering}
\address{Ankara, 06560, Turkey}
\author{N. Kaan Caliskan }
\address{TUBITAK-SAGE}
\address{Advanced Materials Technologies Division}
\address{PK 16, Mamak, Ankara, Turkey}

\begin{abstract}
In this study an automated software model using digital image processing techniques is proposed for extracting the image characteristics and contiguity of liquid phase sintered tungsten heavy alloys. The developed model takes a typical image as input and processes it with no human intervention and provides the corresponding image characteristics and contiguity value. The image processing algorithm includes segmentation, binding point extraction, phase connection, particle count and contiguity estimation stages. For the output, microstructural parameters such as tungsten particle size, amount of tungsten phase and contiguity are determined. The model is implemented by using 6 different scanning electron microscope images of liquid phase sintered 90W-7Ni-3Fe and 93W-4.9Ni-2.1Fe allloys. The results indicate that relative to the manual measurements, the automated model can correctly estimate the contiguity with an error in the vicinity of  5.6\% - 2.9\% for these two alloys. The developed software can easily be adapted to be used for other microstructures. It is also provided as open-source and available for other researchers.
\end{abstract}


\end{frontmatter}


\section{Introduction}
\label{Introduction}
\begin{figure}
   \centering
 \includegraphics[width=.6\textwidth]{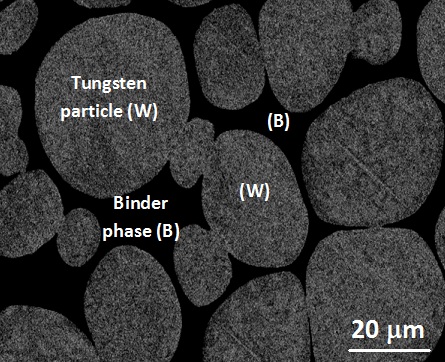}
  \caption{Typical two-phase microstructure of liquid phase sintered tungsten heavy alloy. Bright phase is tungsten (W), and dark phase is binder (B).}
  \label{fig1}
\end{figure}

Tungsten heavy alloys are two-phase metal matrix composites which are generally used in applications such as radiation shields, balance weights and kinetic energy penetrators.  They are manufactured by liquid phase sintering of compacted elemental powders at temperatures above 1450$^{\circ}$C. The two-phase microstructure of liquid phase sintered tungsten heavy alloy consists of a continous network of hard nearly spherical tungsten particles embedded in ductile nickel based binder (or matrix) phase  (Figure \ref{fig1}).

The mechanical properties of two-phase tungsten heavy alloys are controlled by the microstructural parameters such as tungsten particle size, volume fraction of tungsten phase and contiguity, as well as strength of the phases present and strength of the interfaces between them \citep{Muddle1984}. The microstructural parameter contiguity is a quantitative measure of the interphase contact \citep{German1984} and can be defined as the fraction of surface area of a phase shared with grains of the same phase \citep{Gurland1958}. For tungsten heavy alloys, the contiguity C\textsubscript{w} of tungsten network is given by,

\begin{equation}
\label{eq1}
C\textsubscript{w} = 2N\textsubscript{WW} / (2N\textsubscript{WW}  + N\textsubscript{WB} ) 
\end{equation}

where,\newline
N\textsubscript{WW}   -  number of tungsten(W)-tungsten(W) interfaces per unit length of test line,\newline	
N\textsubscript{WB}    - number of tungsten(W)-binder(B) phase interfaces per unit length of test line.\newline

\begin{figure}
   \centering
 \includegraphics[width=.6\textwidth]{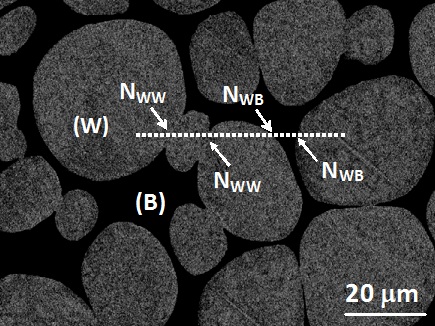}
  \caption{Tungsten-tungsten(WW) and tungsten-binder(WB) interfaces used in contiguity measurements.}
  \label{fig2}
\end{figure}

Equation \ref{eq1} is stated to be valid for any particle size, shape and distribution in two phase composite materials \citep{Aldrich2001}. The values N\textsubscript{WW} and N\textsubscript{WB}  can be measured from the two-phase microstructure of tungsten heavy alloys (Figure \ref{fig2}) by intercept method. This measurement, however, is time consuming. Moreover, large scatter in contiguity data for tungsten heavy alloys had been reported in the literature \citep{Bollina2004,Humail2007,Prabhu2014,Hu2015,Zhou2014,Durlu2014,Eroglu2000,Zhou2009,Hu2015a,Das2010}. Hence, the aim of this study is to use digital image processing techniques in measuring contiguity as well other microstructural parameters such as particle size, and the amount of binder phase. Accurate measurement of these parameters is important in understanding the liquid phase sintering process and development of high performance tungsten heavy alloys.

As motivated by the above analyses, we decided to implement an automated software that takes a tungsten-metal two-phase raw image captured through scanning electron microscope and not only automatically determines the individual tungsten particle characteristics such as size, diameter, circularity, etc. but also calculates the contiguity of the microstructure without any manual intervention. To the best of our knowledge, our study is the first attempt in providing  an automated contiguity estimation of any microstructure in the literature. Furthermore, as far as we know, tungsten heavy alloy two-phase microstructure characterization has not been implemented through automated image analysis techniques in the ternary W-Ni-Fe system. The closest research to our work can be seen in  studies \citep{Shen2005,Kang2000,Tarrago2016}. However, W-Ni-Cu alloy was utilized in \citep{Shen2005}, whereas cemented carbides were investigated in \citep{Kang2000,Tarrago2016}. Moreover, contiguity estimation was not done automatically in these studies. Hence, our primary objectives in this work are:

\begin{itemize}
\item Determine the microstructural characteristics of two-phase tungsten heavy alloys through an automated image analysis algorithm,
\item Calculate contiguity automatically and obtain an acceptable value with a low standard deviation.
\end{itemize}

The structure of the paper is as follows:
After this brief introduction, some of the previous implementations of image processing on microstructures will be covered in Section \ref{Image_Analysis}. The experimental procedure used in this study for obtaining the microstructures of liquid phase sintered tungsten heavy alloys will be given in Section \ref{Experiment}. This is followed by the description of proposed model (Section \ref{Model}) which includes image processing algorithms and the results of image analysis (Section \ref{Image}). In Section \ref{Discussion}, the outcome of the study will be discussed and a detailed analysis will be provided. Finally, we will conclude in Section \ref{Conclusions}.

\section{Image Analysis on Microstructures}
\label{Image_Analysis}

In parallel with the advancements in computer technology, image processing algorithms have started being implemented on various applications in earth sciences. Most of the time, these algorithms are developed to assist the researchers to analyze their observations faster and more accurate. Meanwhile, the interest in automated image analysis applications is rising lately.

Tarquini and Favali developed a Microscopic Information System to segment different rock textures using image analysis algorithms \citep{Tarquini2010}. They implemented a grain segmentation procedure and a region growing algorithm to extract the features and stored the resulting data in a GIS-like database. In another study, the grain boundary motion of microstructured rocks is analyzed in order to observe the influence on the rheology of the rock \citep{Becker2008}.  

Segmentation is the process of dividing the different regions of interest in the image into meaningful parts (such as objects, grains, textures, etc.) \citep{Gonzales2007}. Depending on the complexity of the image and the problem itself, different segmentation techniques can be implemented \citep{Grove2011,Berg2002, Griffin2012, Charles2008, Jungmann2014}. In one study, an adaptive window indicator is adapted for choosing an appropriate threshold level for the segmentation of porous materials \citep{Houston2013}. The authors worked on extracting physical characteristic features in multiphase flow using segmentation and image processing techniques \citep{osgouei2015,ozbayoglu2010, osgouei2010flow,ozbayoglu2012hole,ozbayoglu2011,ozbayoglu2012analysis}. In a similar study , physical characteristics like size, circularity and compactness are extracted for different pressure, temperature conditions of multiphase environments with image analysis techniques \citep{strauss2006} in order to estimate foam stability properties.

After the segmentation process, the statistical features of the microstructures that are analyzed can be calculated. Some automated image analysis models are available for such purpose \citep{Lewis2010, Liu2013}. In some automated models, microstructure shape characteristics are extracted for more accurate evaluation \citep{Jungmann2014, Filho2013}.

Digital image processing techniques have been used in the microstructural analysis of powder size and shape \citep{Nazar1996}, for reconstruction of microstructural volume of liquid phase sintered 83W-11.9Ni-5.1.Fe under microgravity conditions \citep{Tewari2000}, and for determining the coordination number distribution of the same alloy which was liquid phase sintered under normal gravity and microgravity conditions \citep{Tewari1999}.  In another study by Shen et. al., image analysis system has been used in determining the microstructural  parameters such as contiguity, particle size, binder phase, neck size and dihedral angle, of liquid phase sintered 88W-9.6Ni-2.4Cu alloy \citep{Shen2005}. Walte et. al. also used image processing techniques to extract shape characteristics, in particular dihedral angle, in a partially-molten crystalline system \citep{Walte2007}.

Image analysis techniques have also been used in microstructural analysis of cemented carbides. In a study by Kang et. al., optical and scanning electron microscope images of liquid phase sintered cemented carbides were characterized by a commercial automatic image analyzer \citep{Kang2000}. The size, shape and binder phase amount have been correlated to the mechanical properties of cemented carbides. In a recent study by Tarrago et. al., microstructural characterization of cemented carbide samples were done with automated image processing and analysis techniques  \citep{Tarrago2016}. Thresholding, open-close morphological operations, Euclidian Distance Map, Binary Sum and Find Maxima operations were used in order to enhance the image and extract the individual carbide grains. For the estimation of contiguity a double exponential term which includes binder phase amount and mean particle size has been developed. A large scatter in contiguity data was observed due to grain size effect and discrepancies associated with identification and definition of particle boundaries  \citep{Tarrago2016}. These methodologies, in particular, segmentation \citep{Gonzales2007}, thresholding \citep{Gonzales2007}, etc. and algorithms like Watershed \citep{Najman1994}, Hough transform \citep{Duda1972} can also be applied easily to other microstructural  analyses.

\section{Experimental Procedure}
\label{Experiment}

\begin{figure}
\centering
\begin{subfigure}{.32\textwidth}
 \centering
  \includegraphics[width=\textwidth]{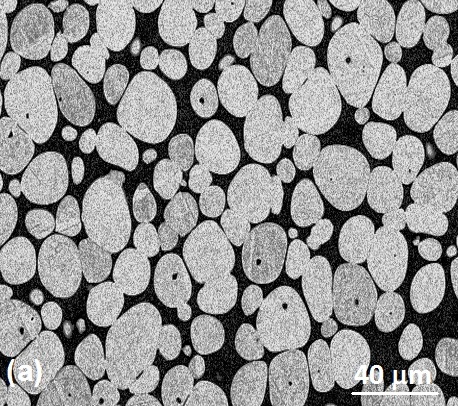}
  \label{fig:sub1}
\end{subfigure}\hspace{0.5em}%
\begin{subfigure}{.32\textwidth}
  \centering
  \includegraphics[width=\textwidth]{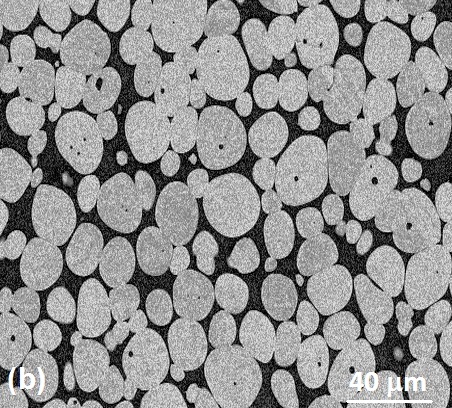}
  \label{fig:sub2}
\end{subfigure}\hspace{0.5em}%
\begin{subfigure}{.32\textwidth}
  \centering
  \includegraphics[width=\textwidth]{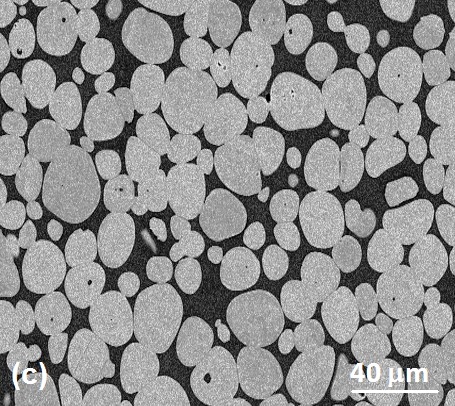}
  \label{fig:sub3}
\end{subfigure}%
\caption{Microstructures of liquid phase sintered 90W-7Ni-3Fe samples utilized in determining the microstructural parameters.}
\label{fig3}
\end{figure}

\begin{figure}
\centering
\begin{subfigure}{.32\textwidth}
 \centering
  \includegraphics[width=\textwidth]{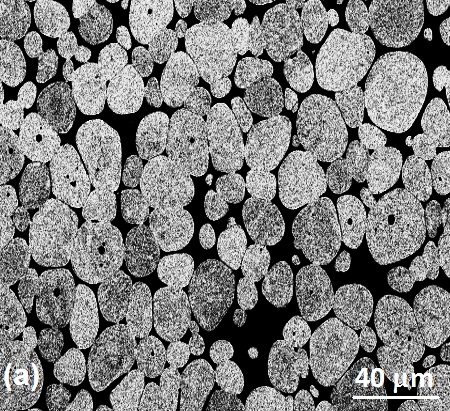}
  \label{fig:sub1}
\end{subfigure}\hspace{0.5em}%
\begin{subfigure}{.32\textwidth}
  \centering
  \includegraphics[width=\textwidth]{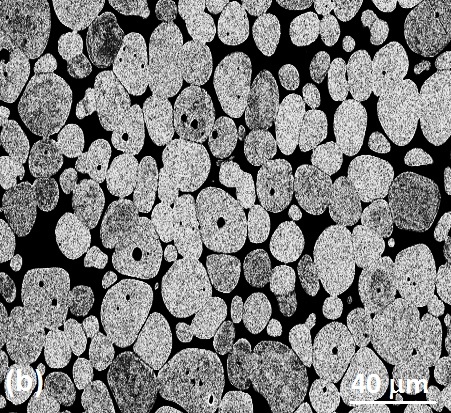}
  \label{fig:sub2}
\end{subfigure}\hspace{0.5em}%
\begin{subfigure}{.32\textwidth}
  \centering
  \includegraphics[width=\textwidth]{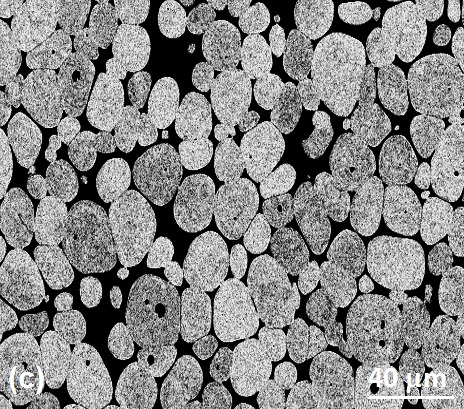}
  \label{fig:sub4}
\end{subfigure}%
\caption{Microstructures of liquid phase sintered 93W-4.9Ni-2.1Fe samples utilized in determining the microstructural parameters.}
\label{fig4}
\end{figure}

Tungsten heavy alloys of the composition (wt.\%) 90W-7Ni-3Fe and 93W-4.9Ni-2.1Fe were prepared by powder metallurgy techniques. Elemental powders of Fe, Ni and W with the appropriate composition were mixed and then cold isostatically pressed under 300 MPa. The green samples were liquid phase sintered at 1480$^{\circ}$C under hydrogen for 20 minutes and then 10 minutes under argon, followed by furnace cooling in argon atmosphere. Microstructural characterization of samples was done with scanning electron microscope (6400 JSM, JEOL Ltd., Japan). In determining the average particle size of tungsten particles, volume fraction of binder phase and contiguity, 6 backscattered scanning electron micrographs obtained at 500x magnification was utilized (Figure \ref{fig3} and \ref{fig4}). Details of experimental procedure was provided in \citep{Durlu2014}.

\section{Proposed Model}
\label{Model}

In order to correctly extract the necessary microstructural parameters and give an accurate contiguity estimation, and more importantly to provide an automated system that will work without any human interaction, the developed algorithm needs to be insensitive to different lighting conditions that the input image is taken. Hence, the proposed model includes the following steps: 

\begin{enumerate}
\item Image retrieval
\item Thresholding (converting to binary)
\item Morphological image processing (noise elimination)
\item Segmentation (extracting the tungsten particles, i.e. the binded tungsten particles are treated together, detecting the tungsten binding locations, connecting the corresponding binding points in such a way that the tungsten particles that are binded together can be counted separately)
\item Image analysis and estimation of particle size, amount of binder phase, contiguity, dihedral angle, etc.
\end{enumerate}

Details of the algorithm are explained in the following subsections:

\subsection{Image Retrieval}
The model takes an image as input. There are no restrictions on how big and what type of image will be presented to the system, but it is assumed that some sort of standardization will be available as such, the image size and type will be set and the user will provide the image accordingly. In this particular study, 6 JPEG images from 2 sets of microgrid structures are used. Each image had 2259 x 1670 pixels resolution and they represented approximately an area of 215 x 159  ${\mu}$$m^{2}$.  


\subsection{Thresholding}

The initial image is in RGB or grayscale format. If it is in RGB, a simple conversion to grayscale is performed, then the image will include pixel values that vary between 0 and 255 \citep{Gonzales2007}. In order to perform the image analysis and extract the useful features, the image needs to be converted to binary. However, depending on the external lighting conditions and the corresponding intensity levels of the image components, this can be a challenging task, since it is not easy, if not impossible, to determine what threshold value should be chosen without further analysis in order to best extract the tungsten particles with minimal information loss. 

\begin{figure}
   \centering
 \includegraphics[width=.8\textwidth]{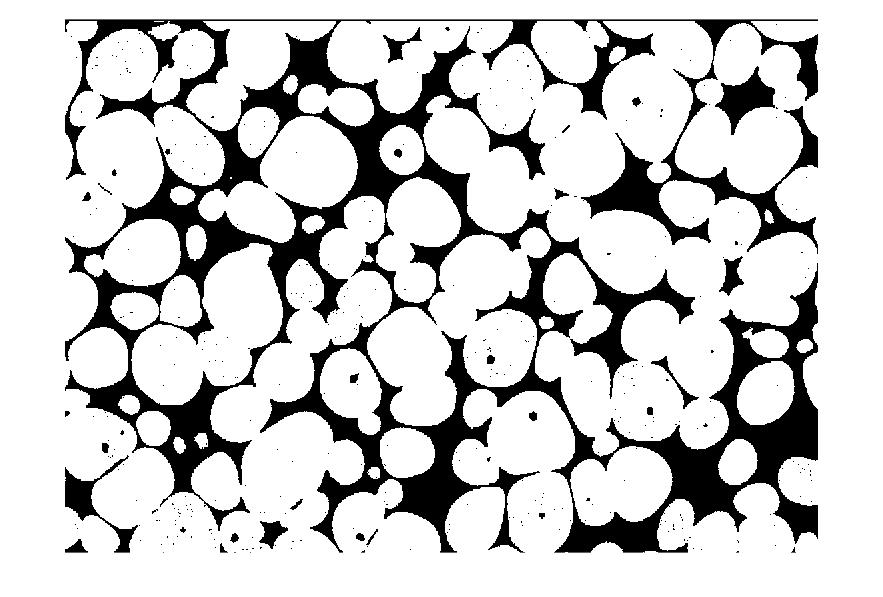}
  \caption{Binary image after thresholding}
  \label{fig6}
\end{figure}

A simple heuristics is used just for that purpose. After going through some extensive analyses of different images, it is observed that the number of very small tungsten particles (i.e. less than 1 ${\mu}$m in size) within the image is very limited (i.e. $<$ 20 for an image size of 215 x 159 ${\mu}$$m^{2}$).  However, if the threshold value is not correctly set, after binarization, a lot of small particles appear in the processed image which indicates the threshold is set high. Through experimentation, it is seen that this value goes down dramatically just when the correct threshold value is chosen. The resulting image after thresholding is shown in Figure \ref{fig6}.

As can be seen from Figure \ref{fig6}, even though the binary image conversion conserves the overall structure of the tungsten particles very well, small amounts of black regions within the particles are still present. These are very small in size, (generally referred as salt and pepper noise) and not to be confused with the small black regions inside the tungsten particles.

\subsection{Morphological Image Processing}
In order to remove the undesired noise from the binary image, a morphological operation is implemented a few times repeatedly. First, for initial cleaning, a simple majority 3x3 filter \citep{Matlab2016} is used where the intensity value of the center pixel of this 3x3 mask is determined by the majority rule in this 3x3 area, i.e. if the majority of the 9 pixels (5 out of 9) within that 3x3 neighborhood is 0 (black), the center pixel will be black, otherwise it will be white. This operation is performed a few times in order to get rid of all the small holes and isolated islands from the image. 


\subsection{Segmentation}

The tungsten particles are ready to be extracted from the metal phase at this stage. The image components (which are the tungsten particles) are extracted from the image and each of them are labeled separately. However due to the tungsten-tungsten binding, most of the particles are connected together and the particles within each component (piece) need to be separated. Algorithm \ref{alg1} takes the image as its input and creates another image in which all the binded tungsten particles can be separately processed. In order to achieve this separation, each connected piece needs to be extracted one by one and processed accordingly. This is achieved by implementing Algorithm \ref{alg2}. Detailed explanations of these algorithms are provided in the Appendix section. 

\begin{figure}
\centering
\begin{subfigure}{.5\textwidth}
  \centering
  \includegraphics[scale=0.33]{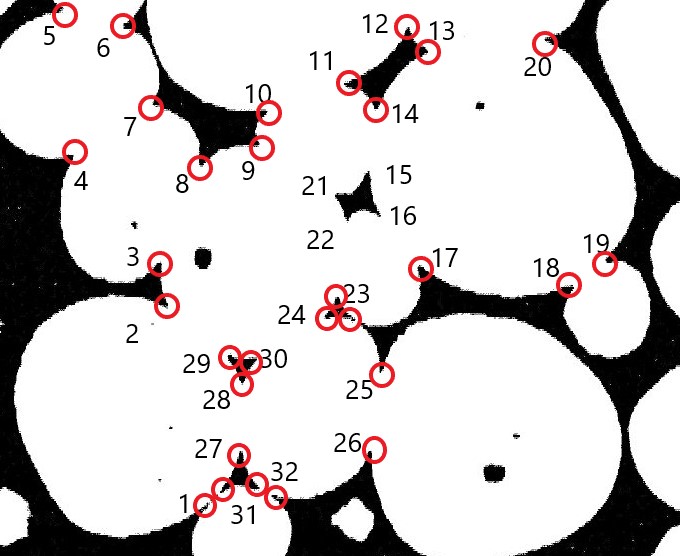}
  \caption{The binding points are identified \\ and extracted.}
  \label{fig10a}
\end{subfigure}%
\begin{subfigure}{.5\textwidth}
  \centering
  \includegraphics[scale=0.33]{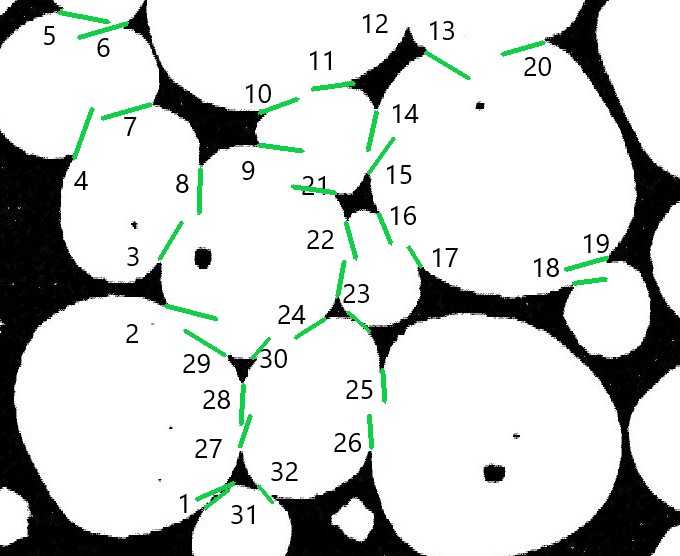}
  \caption{The coordinates and directions of the binding points are determined.}
  \label{fig10b}
\end{subfigure}
\bigskip\par
\begin{subfigure}{.5\textwidth}
   \centering
   \includegraphics[scale=0.33]{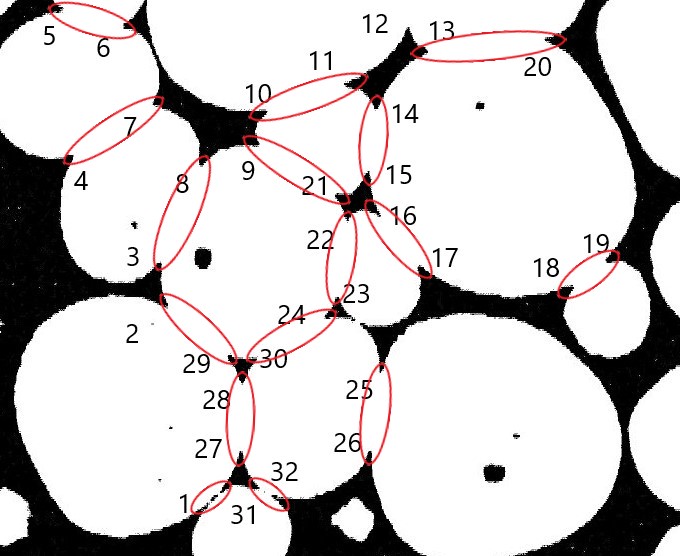}
   \caption{The binding points are matched \\ accordingly.}
   \label{fig10c}
\end{subfigure}%
\begin{subfigure}{.5\textwidth}
\centering
   \includegraphics[scale=0.33]{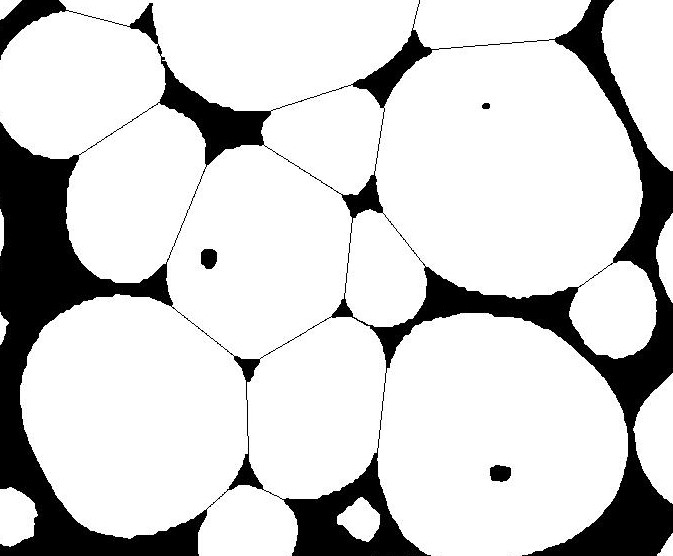}
   \caption{The tunsgten particles are getting separated by connecting the matched pairs.}
   \label{fig10d}
\end{subfigure}
\caption{The process of separating the tungsten pieces that are binded together}
\label{fig10}
\end{figure}

Each connected piece is separated into corresponding tungsten particles using Algorithm \ref{alg2}. Through Figure \ref{fig10}, the results of applying the steps in Algorithm \ref{alg2} can be followed. 

In order to find the binding points, the binding point detection algorithm is used. In that algorithm, the chain code \citep{Matlab2016} is followed. Whenever there is a sudden change in the direction of the chain code (more than 90 degrees) and if there is binding phase in between the two sides of the chain code, in other words, the change in the direction is concave, then that particular point is considered as a binding point.  Figure \ref{fig10a} illustrates the candidate points identified by the binding point detection algorithm.

For every detected binding point, their x-y coordinates and the inward directions are calculated.  The inward direction is calculated by finding the average of the two directions that were acquired from the chain code. The corresponding directions are illustrated in Figure \ref{fig10b}.

As can be seen from the figure, the accuracy of the directions might not be close to the actual value, but, in order to find the matching pairs it should be sufficient, since the pair matching algorithm has some error tolerance for those directional pairing. The algorithm checks the distance between each point and also tries to match their directions. Among the candidate points, the best matching pairs are identified and those are paired.  The result of the pair matching is shown in Figure \ref{fig10c}.

The pair matching algorithm also checks the connecting imaginary line between the two matched pairs. This particular line must pass only through tungsten phase, if there is a mismatch, that particular pair is unmatched and other possibilities are examined. Also, there is a distance threshold for each connection, meaning the binding line cannot be more than a certain length threshold, which means the points which are in close proximity and facing each other have the highest probability of being matched. As can be seen from Figure \ref{fig10c}, it is not necessary to find a matching pair for every point, point number 12 is unmatched. 

At this point, the matching pairs are determined, so we connect these paired points with straight lines. With that, all particles that make up the initial combined piece will be separated into individual tungsten particles so that it will be easier to count the particles. The resulting image after drawing the connecting lines is shown in Figure \ref{fig10d}.

\begin{figure}
   \centering
 \includegraphics[width=.8\textwidth]{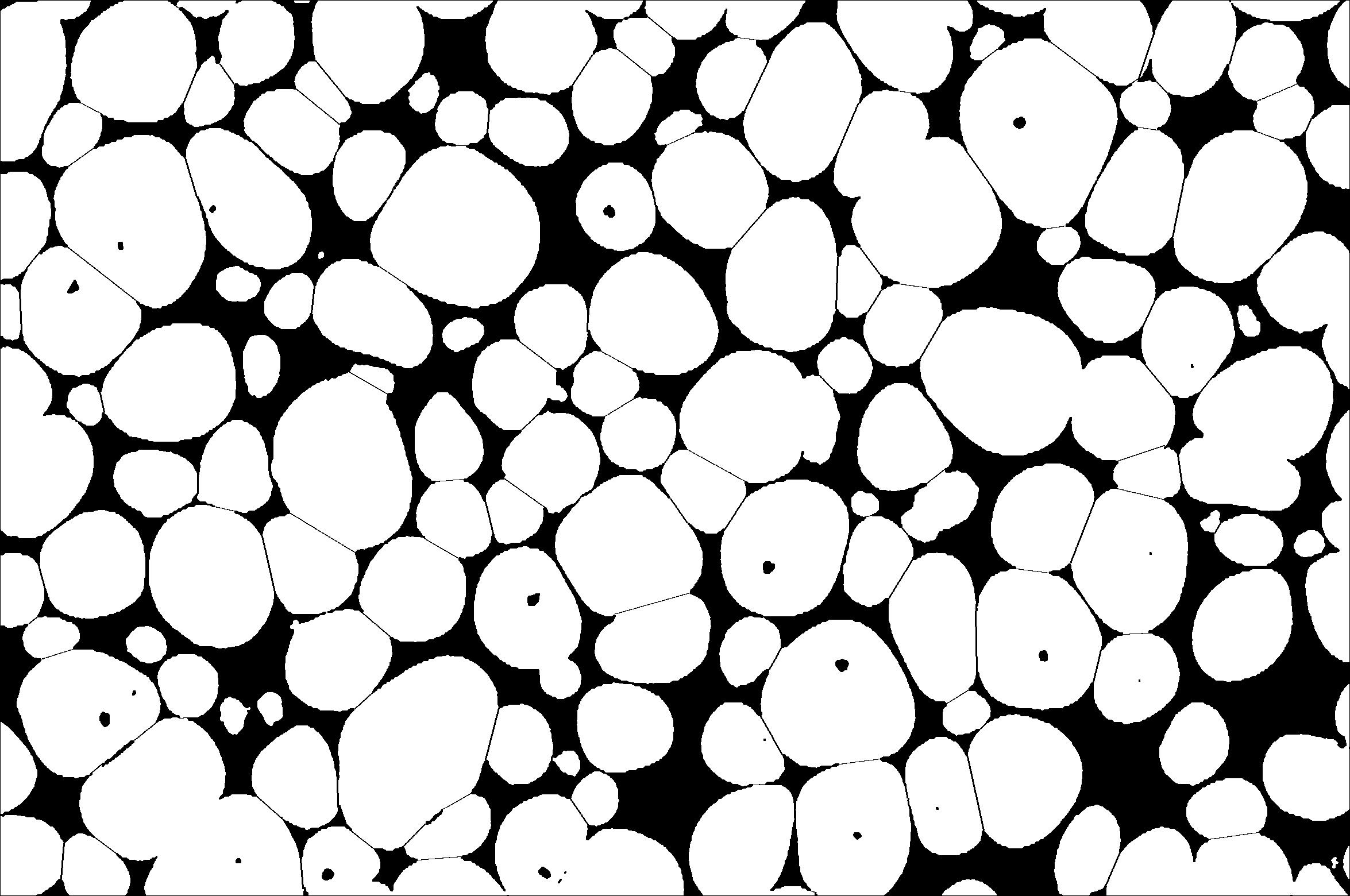}
  \caption{Final image after the particle separation process}
  \label{fig11}
\end{figure}

The final image where the tungsten particles are artificially separated for image analysis is presented in Figure \ref{fig11}. Note the small binder phase regions inside the tungsten particles. Even though some of the connections are either missing or binded wrong, the overall structure is adequate for the image analysis.

\subsection{Contiguity Estimation}

After segmenting the individual tungsten particles separately, the automated contiguity estimation can be achieved. The success of this estimation procedure depends on the performance of the segmentation results obtained through the image processing algorithms.

The details of the contiguity estimation method is as follows: Horizontal and vertical $N_{WW}$  and $N_{WB}$  values are obtained by counting the phase toggles in each line. For  $N_{WB}$, the initial image before particle separation is used so that the binding necks do not exist during the counting process. For  $N_{WW}$, the final image after segmentation is used, so that the difference between the two images (initial and final segmented images) comes directly from the binding connections (necks) between the binded particles.Then these values are used for estimating horizontal and vertical contiguity results by applying Equation \ref{eq1}. 

Meanwhile different mesh sizes varying from the smallest possible grid size (1 pixel wide, which is roughly 0.1 microns) to  10 microns are used during the counting process and the corresponding contiguity values are compared. However no significant variances are observed between the readings.

\section{Image analysis}
\label{Image}

At this point, from the resulting image, it is possible to acquire microstructural parameters. We analyzed 6 different tungsten-metal two-phase microstructures (3 images for 90W-7Ni-3Fe and 3 images for 93W-4.9Ni-2.1Fe) and implemented the proposed automated model on this data. The initial images are 2259x1601 pixel size RGB images stored in JPEG format. Some of the metrics used in our study are provided in Table \ref{table1}.

\begin{table}
\begin{center}
\caption{Metrics used in the developed model} 
\label{table1}
\noindent\begin{tabular}{ |  c | m{5cm} | m{4cm} |  }    \hline
    Used Metric & Description & Formula \\ \hline
    Edge & The edge size of the particle (the particle was assumed square) & $\sqrt{  Area~of~particle  }$ \\ \hline
    Diameter & The diameter of the particle (the particle was assumed circular) & $\sqrt{4 * Area * \pi}$  \\ \hline
    Perimeter & The total pixel (or microns) on the border of the particle & Counted by the algorithm \\ \hline
    Circularity & Indicates how close the particle is to a circular shape & 4 * $\pi$ * Area / $Perimeter^{2}$ \\ \hline
    $N_{WB}$ & Number of tungsten(W)-binder(B) phase interfaces per unit length of test line & Counted by the algorithm \\ \hline
    $N_{WW}$ & Number of tungsten(W)-tungsten(W) interfaces per unit length of test line & Counted by the algorithm \\ \hline
    Contiguity & Fraction of surface area of a phase shared with grains of the same phase & C\textsubscript{w} = 2N\textsubscript{WW} / (2N\textsubscript{WW}  + N\textsubscript{WB} ) \\ \hline
    \end{tabular}
 \end{center}
\end{table}

After implementing the tungsten particle separation, microstructural parameters such as tungsten particle diameter, binder phase amount, and contiguity are extracted from the 6 scanning electron microscope images given in Figures \ref{fig3} and \ref{fig4}. The results are given in Tables \ref{table2a} and \ref{table2b}. In determining the $N_{WW}$, $N_{WB}$ and contiguity values, all the image characteristics, that is, binder phase inside tungsten particles are included into the calculation.

\begin{table}
\begin{center}
\caption{Microstructural parameters of liquid phase sintered 90W-7Ni-3Fe tungsten heavy alloy determined by automatic image analysis with a threshold level of 0.20} \label{table2a}
\noindent\begin{tabular}{ | c  | c  | c | c | c | } 
\hline
\multirow{2}{*}{\parbox{2.3cm}{\centering Microstructural Parameters}} & \multicolumn{3}{ c | }{90W-7Ni-3Fe}  & \multirow{2}{*}{Average}  \\  \cline{2-4} & Fig 3a	& Fig 3b	& Fig 3c &  \\ \hline

Particle diameter (${\mu}$m)  &	17.40 &	16.64 &	16.70 &	  16.91 $ \pm $ 0.42 \\  \hline
Binder phase (\%)	& 22.02 &	21.36 &	22.30 &	21.89 $ \pm $  0.48  \\  \hline
 $N_{WW}$   	& 2.16 &	2.73 &	2.80 &	2.56 $ \pm $   0.35  \\  \hline
$N_{WB}$  	 & 18.53 &	18.19 &	18.28 &	18.33  $ \pm $ 0.18  \\  \hline
Contiguity &	0.189 &	0.231 &	0.234 &	0.218 $ \pm $ 0.025  \\  
\hline
\end{tabular}
\end{center}
\end{table}

\begin{table}
\begin{center}
\caption{Microstructural parameters of liquid phase sintered 93W-4.9Ni-2.1Fe tungsten heavy alloy determined by automatic image analysis with a threshold level of 0.20} \label{table2b}
\noindent\begin{tabular}{ | c  | c  | c | c | c | } 
\hline
\multirow{2}{*}{\parbox{2.3cm}{\centering Microstructural Parameters}} & \multicolumn{3}{ c | }{93W-4.9Ni-2.1Fe}  & \multirow{2}{*}{Average}  \\  \cline{2-4} & Fig 4a	& Fig 4b	& Fig 4c &  \\ \hline

Particle diameter (${\mu}$m)  &	17.48	& 16.49 &	17.41 &	 17.12 $ \pm $ 0.55 \\  \hline
Binder phase (\%)	& 15.31 &	17.61 &	14.90 &	15.94 $ \pm $  1.46  \\  \hline
 $N_{WW}$   	& 3.75 &	3.30	 & 3.77 &	3.61 $ \pm $  0.27   \\  \hline
$N_{WB}$  	 & 15.99 &	16.24 &	15.93 &	16.05 $ \pm $ 0.16  \\  \hline
Contiguity &	0.319 &	0.289 &	0.321 &	0.310 $ \pm $ 0.017  \\  
\hline
\end{tabular}
\end{center}
\end{table}

\section{Discussion}
\label{Discussion}






In the previous sections, the automated contiguity determination algorithm combining image processing techniques with some heuristic procedures is explained in detail. In order to validate the results obtained through this automated model, we have to compare the results with manually determined contiguity, binder phase amount and tungsten particle size. The manual contiguity determination is performed by using Equation \ref{eq1} in two separate measurements in the vertical and horizontal directions of the images given in Figures \ref{fig3} and \ref{fig4}. The average particle diameter of tungsten particles is determined by Jefferson Method \citep{Vander1984}.  In determining the binder phase amount, the corresponding region in the image is selected and then calculated with a commercial image analysis software. The manually determined microstructural parameters are given in Tables  \ref{table3a} and  \ref{table3b}.

\begin{table}
\begin{center}
\caption{Microstructural parameters of liquid phase sintered 90W-7Ni-3Fe tungsten heavy alloy determined by manual image analysis} \label{table3a}
\noindent\begin{tabular}{ | c  | c  | c | c | c | } 
\hline
\multirow{2}{*}{\parbox{2.3cm}{\centering Microstructural Parameters}} & \multicolumn{3}{ c | }{90W-7Ni-3Fe}  & \multirow{2}{*}{Average}  \\  \cline{2-4} & Fig 3a	& Fig 3b	& Fig 3c &  \\ \hline

Particle diameter (${\mu}$m)  &	17.28 &	16.55 &	17.28 &	 17.04 $ \pm $ 0.42 \\  \hline
Binder phase (\%)	& 21.75	& 20.96 & 	24.87	& 22.53 $ \pm $ 2.07  \\  \hline
 $N_{WW}$   	& 2.56	& 3.27 &	2.91	& 2.91 $ \pm $ 0.36  \\  \hline
$N_{WB}$  	 & 18.08 & 	16.68 &	17.62 &	17.46 $ \pm $ 0.71  \\  \hline
Contiguity &	0.221 &	0.282 &	0.248 &	0.250 $ \pm $ 0.030  \\  
\hline
\end{tabular}
\end{center}
\end{table}

\begin{table}
\begin{center}
\caption{Microstructural parameters of liquid phase sintered 93W-4.9Ni-2.1Fe tungsten heavy alloy determined by manual image analysis} \label{table3b}
\noindent\begin{tabular}{ | c  | c  | c | c | c | } 
\hline
\multirow{2}{*}{\parbox{2.3cm}{\centering Microstructural Parameters}} & \multicolumn{3}{ c | }{93W-4.9Ni-2.1Fe}  & \multirow{2}{*}{Average}  \\  \cline{2-4} & Fig 4a	& Fig 4b	& Fig 4c &  \\ \hline

Particle diameter (${\mu}$m)  &	16.78 &	17.03 &	17.47 &	17.09 $ \pm $ 0.35 \\  \hline
Binder phase (\%)	& 16.53	& 19.27 &	13.97	& 16.59 $ \pm $ 2.65  \\  \hline
 $N_{WW}$   	& 4.19	& 4.09 &	4.55 &	4.28 $ \pm $ 0.24   \\  \hline
$N_{WB}$  	 & 15.72 &	17.09 &	15.45 &	16.09 $ \pm $ 0.88  \\  \hline
Contiguity &	0.348 &	0.323 &	0.371 &	0.347 $ \pm $ 0.024  \\  
\hline
\end{tabular}
\end{center}
\end{table}

The results of automated calculation and manual determination of tungsten particle diameter showed very similar results  ($\backsim1$\%) , whereas the difference between two types of measurements for the binder phase amount is less than 4\%.  Hence the developed algorithm is very successful in determining tungsten particle size and binder phase amount in two-phase tungsten heavy alloys.

On the other hand, the difference between automated and manual contiguity measurements is about less than 13\% for both alloys. Such different outcomes for contiguity values indicate that contiguity estimation is not an easy process and can lead to different results depending on the image quality due to sample preparation, examiner (i.e. selected region to be examined) and the chosen model.

At this point, it should be emphasized that there are two different binder phase regions in the microstructure. The first binder phase region is the continuous binder phase which constitutes the majority of the binder phase, the second binder phase region is the one inside tungsten particles and covers only a small portion of the binder phase. The average size of these binder phases is about 1.37 microns (for 90W-7Ni-3Fe alloy) to 1.66  microns (for 93W-4.9Ni-2.1Fe alloy) and they constitute only a very minor amount of the binder phase. The amount of these small binder phases has been calculated with the algorithm as \%0.53 and \%1.18 of the total binder phase for 90W-7Ni-3Fe and 93W-4.9Ni-2.1Fe alloys, respectively.

In the manual measurements of contiguity values a matrix of 10x14 with equal spacings have been used in the horizontal and vertical directions. In order to increase the reliability of the data, two separate contiguity measurements were made and their average value was reported.  In automated contiguity measurements, small amounts of binder phase inside tungsten particles have been included in the calculation of contiguity values. On the other hand, in manual measurements of contiguity, it is a possibility that, small amounts of binder phase inside the tungsten particles, may not be included into the final measured value. In order to elucidate this point, the number of tungsten-binder phase interfaces ($N_{WB}$) and tungsten-tungsten interfaces ($N_{WW}$) have been recalculated with the developed algorithm where binder phase inside tungsten particles have been excluded from the calculation. Accordingly, the number of $N_{WB}$ and $N_{WW}$ interfaces and the corresponding values of contiguity are given in Tables \ref{table4a} and \ref{table4b}.

\begin{table}
\begin{center}
\caption{The calculated number of tungsten-tungsten and tungsten-binder phase interfaces of 90W-7Ni-3Fe tungsten heavy alloy and the corresponding contiguity values with the assumption that there is no binder phase inside the tungsten particles. Threshold level is 0.20.} \label{table4a}
\noindent\begin{tabular}{ | c  | c  | c | c | c | } 
\hline
\multirow{2}{*}{\parbox{2.3cm}{\centering Microstructural Parameters}} & \multicolumn{3}{ c | }{90W-7Ni-3Fe}  & \multirow{2}{*}{Average}  \\  \cline{2-4} & Fig 3a	& Fig 3b	& Fig 3c &  \\ \hline

 $N_{WW}$ (filled)  	& 2.39 &	3.06 &	2.85 &	2.77  $ \pm $ 0.34  \\  \hline
$N_{WB}$  (filled)	 & 18.06 &	17.52 &	18.16 &	17.91 $ \pm $ 0.34 \\  \hline
Contiguity &	0.209 &	0.259 &	0.239 &	0.236 $ \pm $ 0.025  \\  
\hline
\end{tabular}
\end{center}
\end{table}

\begin{table}
\begin{center}
\caption{The calculated number of tungsten-tungsten and tungsten-binder phase interfaces of 93W-4.9Ni-2.1Fe tungsten heavy alloy and the corresponding contiguity values with the assumption that there is no binder phase inside the tungsten particles. Threshold level is 0.20.} \label{table4b}
\noindent\begin{tabular}{ | c  | c  | c | c | c | } 
\hline
\multirow{2}{*}{\parbox{2.3cm}{\centering Microstructural Parameters}} & \multicolumn{3}{ c | }{93W-4.9Ni-2.1Fe}  & \multirow{2}{*}{Average}  \\  \cline{2-4} & Fig 4a	& Fig 4b	& Fig 4c &  \\ \hline

$N_{WW}$ (filled)    	& 4.04 &	3.80 &	3.94 &	3.93 $ \pm $ 0.12   \\  \hline
$N_{WB}$  (filled)  	 & 15.43 &	15.23 &	15.58 &	15.41 $ \pm $ 0.18   \\  \hline
Contiguity &	0.343 &	0.333 &	0.336 &	0.337 $ \pm $ 0.005  \\  
\hline
\end{tabular}
\end{center}
\end{table}

As shown in Tables \ref{table2a}, \ref{table2b}, \ref{table4a}   and \ref{table4b}, exclusion of binder phases inside the tungsten grains led to lower number of tungsten-tungsten interfaces and tungsten-binder interfaces. Such a change led to higher values of contiguity, 0.236, in 90W-7Ni-3Fe alloy, and a   contiguity value of 0.337 in 93W-4.9Ni-2.1Fe alloy. Hence, in manual measurements of contiguity in liquid phase sintered tungsten heavy alloys, it is very likely that due to the exclusion of binder phase inside tungsten particles, relatively higher values of contiguity for tungsten heavy alloys have been reported in the literature.

The difference observed between the automated and manual determination of contiguity can further be explained as follows. The algorithm performance is directly correlated with the image quality. When better images are provided to the system (higher resolution, better lighting conditions and better sample preparation), it is observed that the algorithm performance improves. Another point is that the image processing algorithm depends on finding canditate points from continuous circular flow of the tungsten particle boundary. If such points are found, these are assumed to represent neck intersection points between separate binded particles. However, it is possible that some of these points are not captured by the algorithm, causing the lack of an existing neck being segmented by the algorithm. This causes tungsten-tungsten binding ($N_{WW}$) to go down and this results in a lower contiguity value. Yet another problem might arise due to the existence of small binder phase within the tungsten particles. Most of these binder phase regions are circular in 2-D images and can easily be neglected by the image processing algorithm. However, sometimes their shapes are non-circular and this might cause the algorithm to consider these regions to be possible neck starting points. Hence, non-existent necks can be constructed by the algorithm. Finally, after image processing, some of the tungsten-tungsten bindings might not be detected by the algorithm. The amount of these missing bindings is less than 10\% and it is very likely that in automated calculation of contiguity, due to reduced ($N_{WW}$) values relative to the manual measurements, lower contiguity values were obtained.

\begin{figure}[htp]
   \centering
 \includegraphics[width=.8\textwidth]{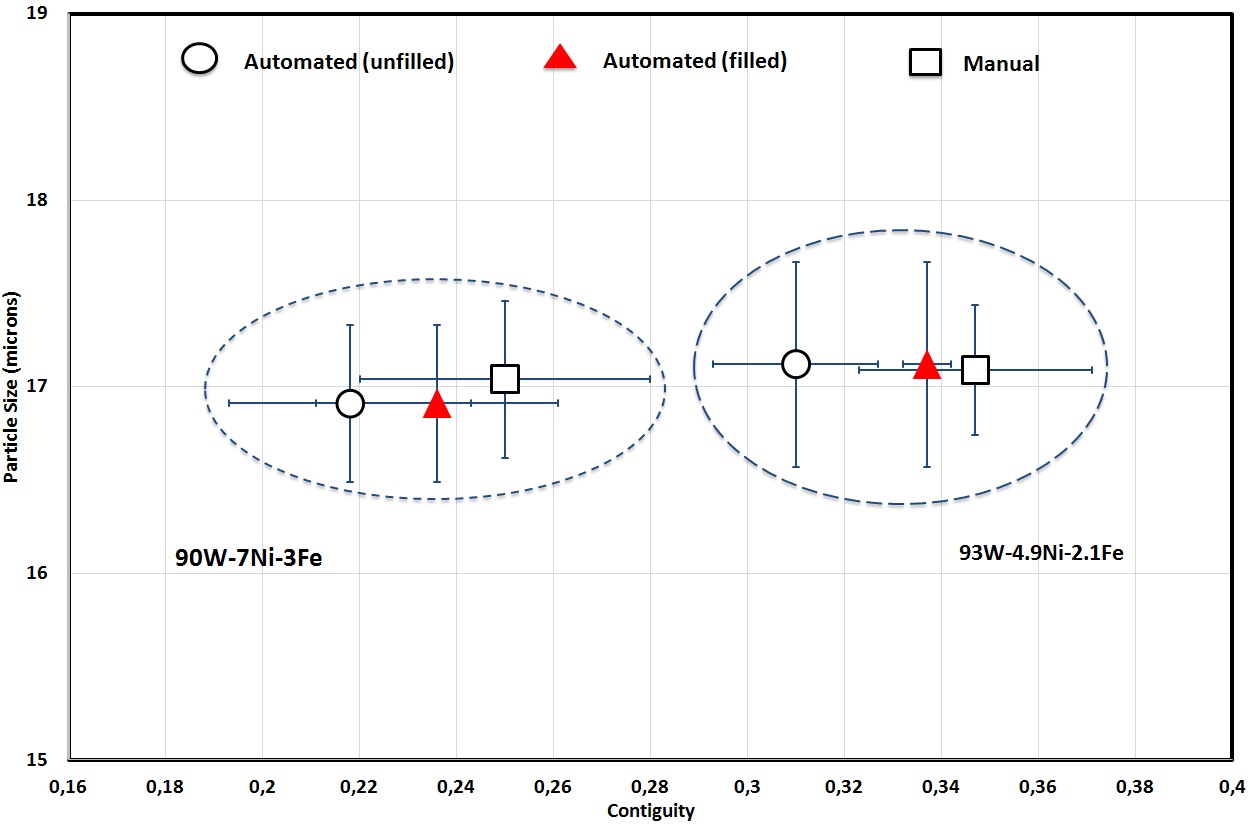}
  \caption{Comparison of automated and manual measurements of  tungsten particle diameter and tungsten contiguity  in liquid phase sintered 90W-7Ni-3Fe and 93W-4.9Ni-2.1Fe alloy}
  \label{fig15}
\end{figure}

\begin{figure}[htp]
   \centering
 \includegraphics[width=.8\textwidth]{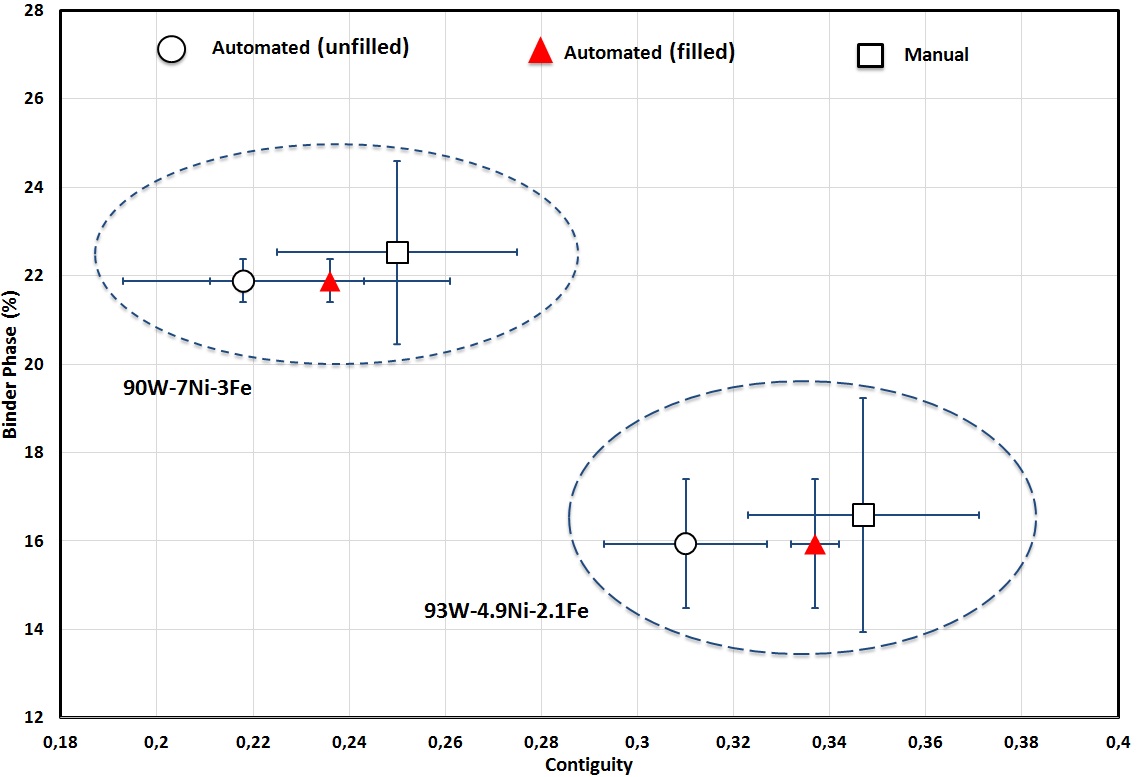}
  \caption{Comparison of automated and manual measurements of  binder phase amount and tungsten contiguity  in liquid phase sintered 90W-7Ni-3Fe and 93W-4.9Ni-2.1Fe alloy}
  \label{fig16}
\end{figure}

We can perform some more analyses by comparing the model parameters and try to come up with some conclusions regarding the relations between them and the contiguity values. Figures \ref{fig15} and \ref{fig16} show the relation between the particle diameter and contiguity, binder phase and contiguity observed through the image processing model, respectively. For the two tungsten heavy alloys investigated, it can be seen from Figure \ref{fig15}  that, there is not a significantly important correlation between the tungsten particle diameter and tungsten contiguity observed through the images.

When we analyze the amount of binder phase against the contiguity values of tungsten phase, it can be seen from Figure \ref{fig16}  that there is a very strong relation between these two microstructural parameters. As the amount of binder phase is decreased (or the amount of tungsten phase is increased), the number of tungsten-tungsten interfaces increases and this leads to higher values of contiguity. 

Since this is a preliminary work, in this study we only used 6 microgrid structure images for developing a contiguity estimation model based on automated image processing. However, even with the limited amount of data, we successfully provided an estimation model for contiguity. Moreover, the developed algorithm also determines other microstructural parameters such as neck size, dihedral angle, circularity and their distributions.  Quick and accurate determination of these parameters will be very helpful in understanding the liquid phase sintering mechanism and their effect on mechanical properties of tungsten heavy alloys. In the near future we are planning to test our system with more images, aiming to reduce the error in the estimation. Meanwhile, since the error is within the acceptable vicinity of variation between different manual readings, this model can be used as a fast contiguity estimation for two-phase liquid phase sintered tungsten heavy alloys, as well as other microstructural parameters.

\section{Conclusions}
\label{Conclusions}

In this study, an automated image analysis system for determining contiguity, tungsten particle size and the amount of binder phase of tungsten two-phase metal matrix composites is proposed. The model is developed by using 6 different scanning electron microscope images of liquid phase sintered 90W-7Ni-3Fe and 93W-4.9Ni-2.1Fe allloys. The results indicate that relative to the manual measurements, the automated model can determine the contiguity of tungsten particles with 14\% error for these two alloys. However, when the binder phase inside the tungsten particles is excluded from the analysis, the contiguity estimation difference between manual and automated calculation is reduced to 5.6\% and 2.9\% respectively. The model  also determines other microstructural parameters such as tungsten particle size with nearly 1\% difference and binder phase amount with 4\% difference.

\section*{Acknowledgements}

This work was supported by TOBB University of Economics and Technology and TUBITAK-SAGE.

\section*{References}



\bibliographystyle{elsarticle-num} 

\appendix
\section*{Appendix}
\label{Appendix}

\begin{algorithm}
\SetAlgoNoLine
\KwIn{$A$ (image)}
\KwOut{Final Image}
{Let $n$ be the number of separate pieces in A (can be single or binded multiple tungsten particles)} \\
{Let $P$ be the collection of these separate pieces, where  $P_{i}$ indicates $i^{th}$ piece.} \\
\For{$i\leftarrow 1$ \KwTo $n$}{
Calculate the Circularity of $P_{i}$ \\
\eIf{$P_{i}$ $>$ Circularity Threshold } {
$P_{i}$ is a single particle, no need to perform any processing.}
{$P_{i}$ consists of multiple particles binded together. \\
We need to extract each particle from the combined piece. \\
Segmented $i^{th}$  Piece $\leftarrow $ Particle Segmentation Algorithm ($P_{i}$)}
}
Final image  $\leftarrow $ combination of all segmented pieces\\
\Return (Final image)
\caption{Process Tungsten Particles}
\label{alg1}
\end{algorithm}

\begin{algorithm}
\SetAlgoNoLine
\KwIn{$CP$ (combined piece)}
\KwOut{$SP$  (separated piece)}
\For {Each hole within $CP$}{
\eIf{hole size $<$ hole size threshold $and$ hole circularity $>$ Hole circularity threshold } {
Fill hole							----- See (i)}
{Return ($CP$)						----- See (ii)}
}
{$cc$ $\leftarrow $ chain code ($CP$)							----- See (iiii))} \\
{$BP$ $\leftarrow $ Binding point detection ($cc$)					----- See (iv)} \\
{Let $n$ be the number of detected binding points.} \\
\uIf{$n$ = 0}
{This is a single particle, no need for further processing} 
\uElseIf{$n$ = 1}
{There is just 1 corner, no need for further processing		----- See (v) }
\ElseIf{$n$ $>$ 1}
{ {$MP$ = find matching pairs ($BP$)				----- See (vi)} \\
$SP$ = draw binding lines ($CP,MP$)					----- See (vii)}
\Return ($SP$) \\


 \renewcommand{\theenumi}{\roman{enumi}}
\begin{enumerate}
\item This is a circular (or round shaped) hole which is inside a single particle. It does not get involved in the tungsten-tungsten binding process, so this hole will not be considered during the binding corner detection.
\item This hole is most probably between two or more tungsten particles, so it will be used during the binding corner detection.
\item Find the chain code of the combined pieces perimeter starting from the most upper left perimeter pixel. Also, if there are holes within the combined piece, get their perimeter chain codes in the same order.
\item $BP$ consists of the coordinates and the direction of each binding point. Using this chain code information, find all possible binding points according to binding point detection algorithm. 
\item We can not join it with another corner, it is most probably one of those points which is close to the edge of the picture frame within a partial particle, we also cannot do anything with it, no need for further processing.
\item Find matching pairs with these binding points according to the binding point pair matching algorithm.
\item Draw lines within the combined piece between the matched pairs, so the tungsten particles will be separated, $SP$ = separated piece. 
\end{enumerate}
\caption{Particle Segmentation Algorithm }
\label{alg2}

\end{algorithm}

\end{document}